\documentclass[11pt, dvips]{elsart}
\usepackage{graphicx}
\usepackage{amssymb}
\journal{Astroparticle Physics}

\newcommand{\DC}{D\v{C} }
\newcommand{\cer}{\v{C}erenkov }
\newcommand{\ie}{\emph{i.e.}}
\newcommand{\eg}{\emph{e.g.}}
\begin{document}
\begin{frontmatter}
\title{A High Resolution Method for Measuring Cosmic Ray Composition beyond 10 TeV}
\author[Utah]{D. B. Kieda\corauthref{CA}},
\ead{kieda@krusty.physics.utah.edu}
\author[Chicago]{S. P. Swordy},
\author[Chicago]{S. P. Wakely}
\address[Utah] {Department of Physics, University of Utah, Salt Lake City, UT 84112, U.S.A.}
\address[Chicago]{Enrico Fermi Institute and Department of Physics, University of Chicago, Chicago IL 60637, U.S.A.}

\corauth[CA]{Corresponding author.}

\begin{abstract}
The accurate determination of the elemental composition of cosmic
rays at high energies is expected to provide crucial clues on the
origin of these particles. Previous direct measurements of
composition have been limited by experiment collecting power,
resulting in marginal statistics above $10^{14}$ eV, precisely the
region where the ``knee'' of the cosmic-ray energy spectrum is
starting to develop. In contrast, indirect measurements using
extensive air showers can produce sufficient statistics in this
region but generate elemental measurements which have relatively
large uncertainties. Here we discuss a technique which has become
possible through the use of modern ground-based \cer imaging
detectors. We combine a measurement of the \cer light produced by
the incoming cosmic-ray nucleus in the upper atmosphere with an
estimate of the total nucleus energy produced by the extensive
air shower initiated when the particle interacts deeper in the
atmosphere. The emission regions prior to and after the first
hadronic interaction can be separated by an imaging \cer system
with sufficient angular and temporal resolution.  Monte Carlo
simulations indicate an expected charge resolution of $\Delta Z/Z
<5\%$ for incident iron nuclei in the region of the ``knee'' of
the cosmic-ray energy spectrum. This technique also has the
intriguing possibility to unambiguously discover nuclei heavier
than iron at energies above 10$^{14}$ eV. The identification and
rejection of background produced by charged particles in ground
based gamma-ray telescopes is also discussed.

\end{abstract}
\end{frontmatter}

\section{Introduction}

The origin of cosmic rays remains a central unresolved question in
astrophysics. Nearly 90 years after the first observations, this
population of charged particles remains an enigma; the heart of
which is the huge dynamic range of fluxes and energies over which
they have been observed. In the theoretical arena, a relatively
recent paradigm has emerged, which involves the acceleration of
cosmic rays via diffusive shock processes in supernovae remnants
(SNR). This idea has been fueled both by a viable physical model
of the acceleration process
\cite{Bell:1978a,Bell:1978b,Blandford:1978,Krymsky:1977} and a
simple energetics argument in which galactic supernovae are the
only galactic candidate with sufficient energy output to supply
the staggering amount of power needed to sustain the cosmic-ray
population (see \eg, \cite{Schlickeiser:1994}).

A key issue with the SNR idea is that supernova diffusive shock
acceleration can only produce particles up to some maximum energy,
limited either by the lifetime of the strong shock or by the
particles becoming so energetic they can no longer be confined in
the acceleration region \cite{Lagage:1983}. Estimates of this
upper energy limit vary, with typical values in the region of
$10^{14}$ eV.  However, the observed flux of cosmic rays extends
more or less continuously for another five orders of magnitude
beyond this; no plausible extrapolation of the standard parameters
of SNR can generate cosmic rays of such energies.

Another aspect of the cosmic ray riddle is the existence of an
observational ``knee'' at $\sim 10^{15}$ eV in the cosmic ray
energy spectrum. The coincidence of this ``knee'' with the
theoretical energy limit of SNR diffusive shock acceleration is
intriguing and has actually served as evidence to support the
theory. However the ``knee'' represents only a small change in the
spectral slope of the overall flux, with the energy dependence
changing from $E^{-2.75}$ below the ``knee'' to $E^{-3.0}$ above
it.  We are faced with a simple observational fact that the cosmic
rays have an essentially continuous spectral slope for nearly 11
orders of magnitude.

In principle, additional mechanisms could provide the flux at high
energies.  The power budget for ``post-knee'' cosmic rays is only
a fraction that of the total budget, so there is some freedom in
selecting models.  However, to result in an energy spectrum as
smooth as is observed, these mechanisms would have to generate
fluxes which are remarkably, perhaps implausibly, close to that of
the SNR mechanism. Another problem exists at the highest energies,
around $\sim 10^{19}$ eV, where an additional spectral break occurs (the
``ankle'').  It has been argued that particles above the
``ankle'' could be extragalactic.  If so, they can be expected to
exhibit a characteristic cutoff due to photo-pion production with
the cosmic microwave background.  This cutoff has yet to be
observed \cite{Hayashida:2000}.

An overall scheme which credibly addresses and unifies these
issues remains elusive.  More reliable and accurate measurements
in the ``knee'' region, of abundant cosmic ray  nuclei (1 $\le Z
\le$ 26)  could drastically revise our current ideas and provide a
basis for such a unification. The accurate determination of the
composition of cosmic rays has provided some of the key advances
in this field at lower energies, where direct measurements are
possible with detectors above the atmosphere. For example, the
realization that the observed spectral slope of cosmic rays is
significantly steeper than that produced in the cosmic ray
sources themselves resulted from measurements with sufficient
elemental resolution to separate primary source cosmic ray
elements from those produced in the interstellar medium at
$10^{11}$ eV \cite{Juliusson:1972}.

In this paper we propose an idea which has become possible
through advances in the imaging atmospheric \cer technique.  It is
based on the concept that a detector of fine enough pixelation
will be capable of observing \cer light emitted directly from
cosmic ray nuclei prior to their first interaction in the
atmosphere.  In general, this light is overwhelmed by the \cer
emission from the subsequent extensive air shower (EAS).  However,
with an appropriate detector, and within certain geometric
constraints, this ``direct \v{C}erenkov", or D\v{C}, light can be
sufficiently well separated from the background of EAS \cer light
to make relatively high-precision measurements of the traditional
cosmic ray composition ($1 \le Z \le 26$). This technique may
also provide new opportunities for the first measurement of
higher charge nuclei (Z $\gg 26$) as well as possible
improvements in ground based gamma-ray telescope sensitivity.

\section{Method\label{sec:Method}}

\subsection{History}

\cer light from extensive air showers was first predicted by
Blackett in 1948 \cite{Blackett}, and later observed by Galbraith
and Jelley in 1952 \cite{GalbraithJelley}. In 1989, the Whipple
telescope used the technique of \cer light imaging to provide the
first highly significant detection of the Crab Nebula in high
energy gamma rays \cite{Weekes89}. The imaging technique, which
was proposed in 1977 \cite{WeekesTurver}, involves the use of an
array of photomultiplier tubes at the focal plane of the
telescope to reject the very large background of cosmic
ray-induced air showers by the shape of their images in the field
of view. A camera with fine enough pixelation has the capacity to
not only discriminate between hadronic and electromagnetic
showers, but to map the development of the showers as they
penetrate into the atmosphere.

In the present work we propose a means to measure the direct \cer
(D\v{C}) light produced by the incoming nucleus prior to its first
interaction by using the imaging atmospheric \cer techniques of
VHE gamma ray astronomy. The targeting of D\v{C} light is not a
new idea. In 1965, Sitte \cite{Sitte} had proposed that this
radiation might be observed in high-altitude balloon-borne
instruments. His idea, which was revisited by Gough in 1976
\cite{Gough}, was to place detectors at a height in the
atmosphere above the mean interaction point of heavy primary
cosmic rays and to look for the direct production of \cer light.
In the absence of the large light backgrounds due to emission from
EAS, it was expected that accurate composition measurements could
be made by analysis of the direct \cer light yields. After a
pioneering flight by Sood in 1981 \cite{Sood83}, the concept was
unexploited until recent efforts by Seckel \emph{et al.} in 1998
\cite{BACH99}. The fundamental challenge in the new approach
discussed here is the identification of the D\v{C} light against the
much larger background of \cer light produced in the associated
EAS.

\subsection{\cer Radiation}

A charged particle traveling in the atmosphere will produce \cer
radiation if it has a velocity greater than the local velocity of
light. The threshold Lorentz factor, $\gamma_0$, at which the
radiation starts to be emitted is approximately:
\begin{equation}
  \label{eq:CerAngle}
   \gamma_0 \approx \frac{1}{\sqrt {2 \delta}}
\end{equation}
where $\delta = n-1$, and $n$ is the local index of refraction in
the atmosphere. At sea level, $\gamma_0\approx 42$, while at an
altitude of some 50 km, $\gamma_0\approx 680$.  For an Iron
nucleus, this is equivalent to energies of 2 TeV and 36 TeV,
respectively.

The rate of emission of \cer light, $N_{\check{C}}$, increases
rapidly with $\gamma$ above this threshold as:
\begin{equation}
\label{eq:CerEm}
 N_{\check{C}} \propto Z^2 \bigl( \frac{1}{\gamma_0^2}-\frac{1}{\gamma^2}
 \bigr)
 \end{equation}
where $Z$ is the particle charge. For $\gamma \gg \gamma_0$ the
amount of emitted light approaches a saturation level which
scales exclusively with the square of the particle charge and the
local atmospheric density. The Lorentz threshold, $\gamma_0$, is a
function of the local density, which, in turn, is a function of
the altitude.  Therefore, because the atmospheric density scales
with height, once in saturation, the amount of \cer emission from
a particle is determined entirely by the charge of the particle,
and the altitude of emission.  The \cer angle too, is determined
by the local density (and thus, on the emission height), and so
the light pool from an emitting particle has a well-defined
geometry determined entirely by the atmospheric density profile.

\begin{figure}
  \centering
  \includegraphics[height=4.0in]{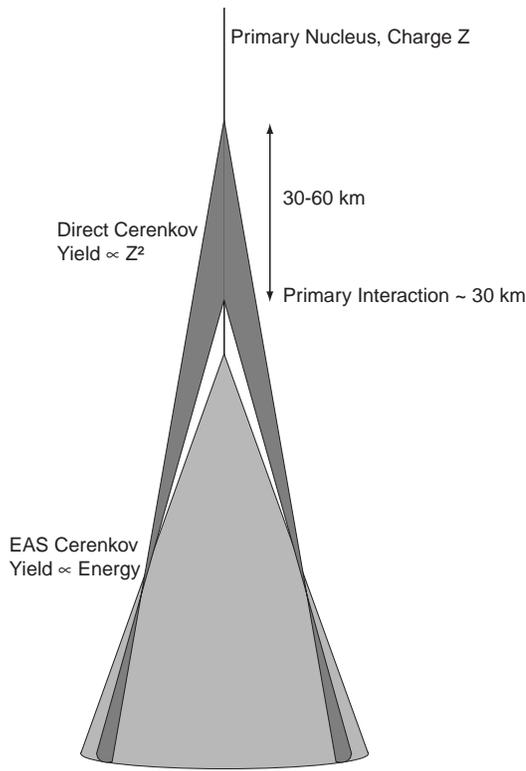}
  \caption{Schematic Representation of the \cer emission from a
           cosmic ray primary.}
  \label{Cones}
\end{figure}

In contrast, the \cer emission from an extensive air shower has a
yield almost linearly proportional to the energy of the primary
particle,  with a weak dependence on the primary charge, and a geometry
which varies primarily with the details of the shower
development. Figure \ref{Cones} shows a schematic depiction of
the \cer radiation from a primary cosmic ray and its subsequent
EAS. In the interior regions of the \cer light
pool, the typical photon densities due to the EAS emission are
many times greater than those associated with the direct
emission.  Thus, to discern the D\v{C} photons against this
background, some distinguishing characteristics must be
identified.

\subsubsection{Angular Characteristics}

One of the fundamental characteristics of atmospheric D\v{C}
light is a simple correspondence between emission height and
emission angle. As shown above, this stems from the density
dependence of the \cer emission angle and the height dependence
of the atmospheric density.  In practice, this means that for
fixed geometries, single emitting particles can be matched by
their emission angle to unique emission heights.

Figure \ref{fig:Lines} shows the characteristic features of
D\v{C} light from a relativistic ($\gamma \gg \gamma_0$) charged
particle vertically incident on the top of the atmosphere.  To
illuminate the features of the D\v{C} emission, the particle is
constrained to not interact in the atmosphere; the intersection
point of the particle trajectory with the observing plane is the
point where the impact parameter is zero.  The default CORSIKA
Monte Carlo simulation atmospheric profile\cite{CORSIKA} is used.

The upper panel shows the impact parameter of photons at sea
level produced as a function of the particle altitude in the
atmosphere. The physics of atmospheric \cer emission produces a
narrow emission cone angle at high altitude which expands as the
particle penetrates deeper into the atmosphere.  At altitudes
around 10 km, the interplay between increasing emission angle and
decreasing emission height produce a pileup effect near the
maximum geometrically-allowed radius, near $\sim 145$ m.  This
pileup is responsible for the so-called ``\cer ring".

Since the rate of emission of \cer light increases with
atmospheric depth (because $\gamma_0$ becomes smaller), the
optimal impact parameter for viewing D\v{C}, in the absence of an
EAS, would be at this \cer ring.  However, since most nuclei, in
reality, interact by altitudes of 25 km or so, the best impact
parameter must be chosen as a compromise between high D\v{C}
density (low altitude), and low nuclei interaction probability
(high altitude).  A reasonable radius is $\sim 80$ m.  At this
radius, a 10 m diameter detector will view \DC light coming from
an angular range of $\sim 0.15$ degrees with respect to the
incoming trajectory, which corresponds to heights between roughly
29 and 34 km in the atmosphere. On average, this is about as deep as the
cosmic ray primary can be expected to penetrate without interaction.

As Figure \ref{fig:Lines} shows, an additional component of secondary \cer light
which originates from emission at an altitude $\sim$ 4 km could
also be observed at 80 m.  However, this light is emitted at an
angle of nearly 1 degree, which is distinguishable from the light
produced at higher altitudes by using an imaging detector with
fine enough pixels. In a telescope with 0.05 degree pixels, the
\DC light from high altitude would be confined to a specific range
of pixels. In the absence of EAS particles, measurements of
photon densities in these angular bins would provide an
unambiguous determination of the \DC yield from the primary
particle, and therefore its charge.

However, the presence of the EAS acts to obscure the direct \cer
component, as electrons produced at lower altitudes get scattered,
producing \cer light which enters into the angular bin of the
direct radiation. In lower energy showers, the effect is small
compared to the densities of the D\v{C} light. However, as the
size of the EAS increases, the background grows to the point
where the density of the scattered electron emission exceeds that
of the D\v{C} emission. Because the amplitude of the EAS light is
set by the energy of the incoming nucleus, this background
effectively provides an upper limit to the useful energy range
for the D\v{C} technique.  This energy limit essentially scales
with the square of the primary nuclear charge.

\subsubsection{Temporal Characteristics}

\begin{figure}
  \centering
  \includegraphics[height=5.0in]{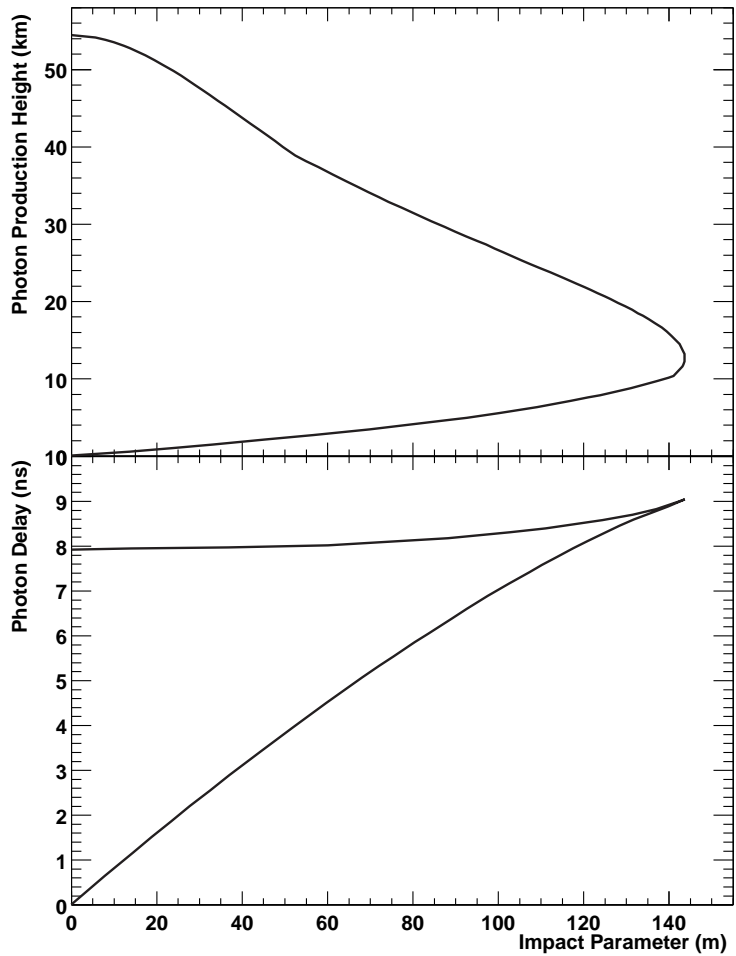}
  \caption{The direct \cer emission characteristics of a single non-interacting particle
   vertically incident on the atmosphere observed at sea level.
   Upper Panel: The \cer photon emission height versus photon impact parameter
		 relative to the original particle trajectory.
   Lower Panel: The photon time delay measured at the observation depth relative to the particle traveling
                 at speed $c$ versus photon impact parameter relative to the original particle trajectory}
     \label{fig:Lines}
\end{figure}

The lower panel of Figure \ref{fig:Lines} shows the time delay of
\cer photons emitted by the same particle as seen in the first
panel. The delays are shown as a function of the photons' final
impact parameters at sea level. These are measured with respect
to the time it takes the particle (assumed to be traveling at
$c$) to reach sea level.

The D\v{C} light has a very fast time structure at high altitude.
All of the photons emitted from altitudes higher than 30 km arrive
within roughly 300 ps of each other. Furthermore, this light is
delayed with respect to the light emitted lower in the atmosphere
(\eg, below 5 km) by 3 ns or more, depending on the observation
radius. The effect of optical dispersion over a typical
wavelength range for atmospheric \cer measurements ($300-600$ nm)
has a negligible effect on these timing widths. This leads to a
additional method for differentiating between the D\v{C} light and that
produced in the EAS. The D\v{C} should be delayed from the main
\cer pulse, and should have a characteristic width over an order
of magnitude shorter than the EAS \cer pulse.  The width of an EAS
\cer pulse depends on the width of the shower front. These fronts
are typically $\approx 2$ m thick, leading to EAS \cer pulses on
order 6 ns wide.

The largest time separation between D\v{C} and EAS \cer light
occurs closest to the core of the shower. However the total
amount of D\v{C} radiation collected is smaller closer to the
core as discussed above. Again, an observation radius of $\sim
80$ m is a good compromise between providing an adequate D\v{C}
signal before the particle is likely to interact and maintaining
some time separation between the D\v{C} and EAS signal as shown in
the lower panel of Figure \ref{fig:Lines}.

\section{Simulation\label{sec:Sim}}
Since the dominant background for the D\v{C} light is \cer light
from electrons scattered in the EAS development, we have used
numerical simulations to study the levels and fluctuations in this
background. The characteristics of D\v{C} light in EAS have been
modeled using a modified MOCCA Monte Carlo Simulation
\cite{Hillas82a,Hillas82b,Kieda95} and also the CORSIKA (Version
5.945, QGSJet98) simulation package \cite{CORSIKA}. Both of these
codes demonstrate similar characteristics for the D\v{C} light,
and predict that D\v{C} light should be observable against the
background EAS \cer light over an energy window which depends on
the charge of the primary particle.

\begin{figure}
 \centering
 \includegraphics[width=5.0in]{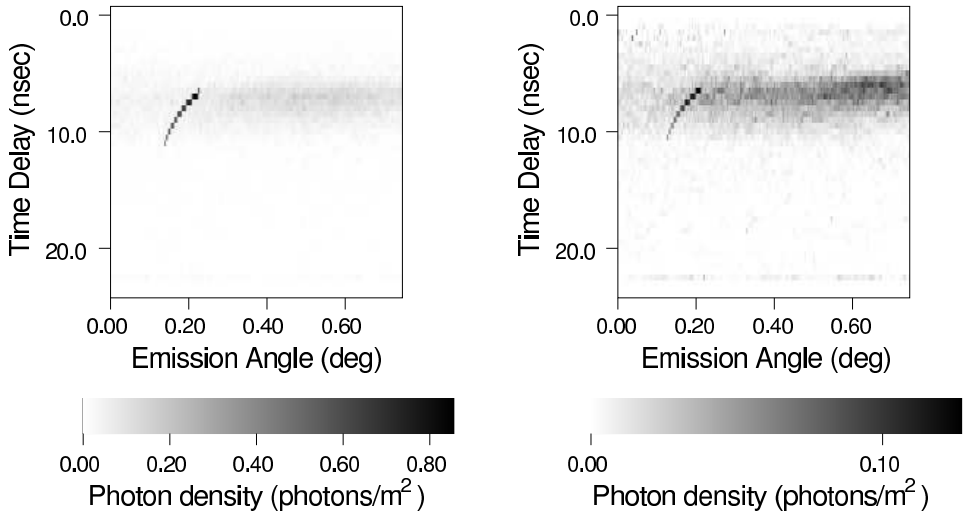}
  \caption{Simulated D\v{C} and EAS \cer light emitted from a single vertically incident
  particle. Left-hand panel shows a 10 TeV $Z=12$ nucleus and
  the right hand panel shows a 5 TeV $Z=4$ nucleus.
  The \cer light is observed at a radius between $67-94$ m (mean radius 80 m) from the
  shower axis.  The vertical axis is the time delay of the arriving
  photons as discussed in the text. The horizontal axis is the
  arrival angle of the photons with respect to the vertical at the observing
  site. The scales below each panel give the photon intensities.}
  \label{new2}
\end{figure}

\subsection{Average Behavior}

First we consider the average behavior of the D\v{C} light signal
from various incoming particles observed at sea level. The left
hand panel of Figure \ref{new2} illustrates the angular and time
characteristics of D\v{C} light emitted by a single 10 TeV
vertical cosmic ray with $Z=12$ (Mg) which interacts and produces
an EAS deeper in the atmosphere. The time axis corresponds to the
time delay compared to a reference arrival time at sea level of a
particle traveling at a speed $c$ along the incoming nucleus
path. The angular axis corresponds to the angle of entry of the
photons into the detector compared to the incoming particle
trajectory. The photon density of the light is averaged over an
annulus extending from 67 m to 94 m from the shower axis, giving an
 80m mean radius of the annulus, and
the intensity is integrated over a wavelength band of $300-600$
nm. The D\v{C} light emission is clearly seen as an arc on the
left of the figure, separated from the \cer emission produced by
the EAS. The right hand panel of Figure \ref{new2} shows the
signal from a single 5 TeV $Z=4$ (Be) nucleus. This shower also
has an obvious D\v{C} signal, but the intensity of this feature
is clearly less well separated from the background.

\begin{figure}
  \centering
  \includegraphics[height=4.0in]{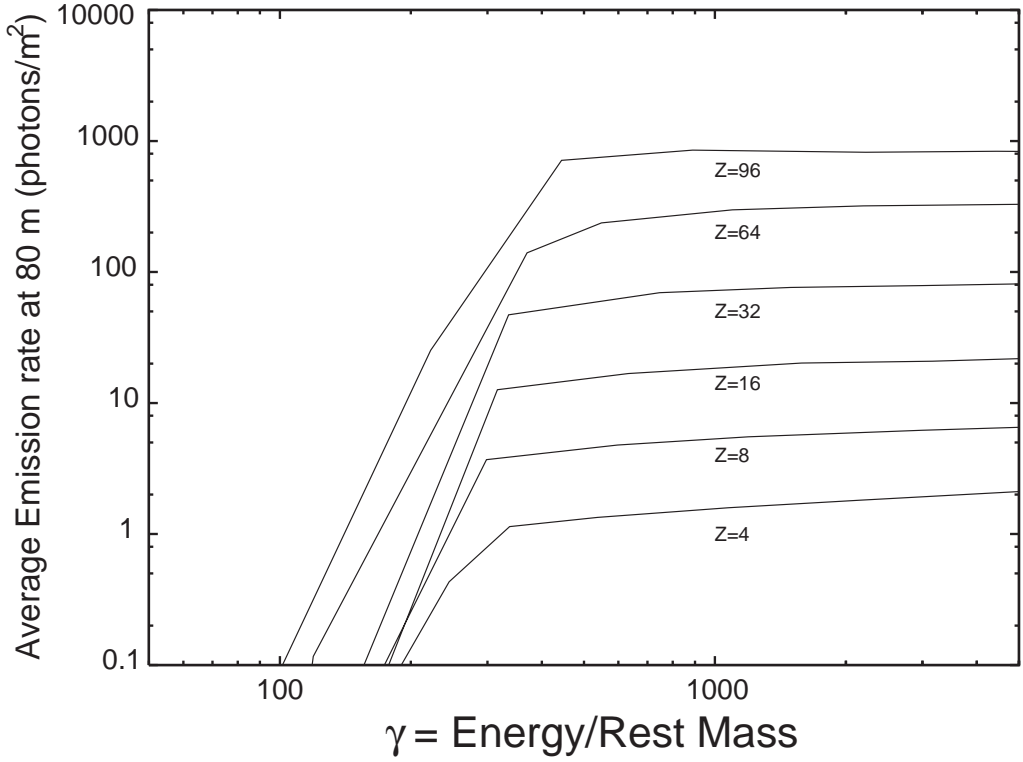}
  \caption{Simulated Average D\v{C} Emission rate for different charge
           primary nuclei. All shower trajectories are vertical and
           measurements are made at sea level.  Vertical Axis:
           Average Emission rate (Photons/m$^2$ at 80 m mean radius from shower core).
       Horizontal Axis: Primary cosmic ray Lorentz factor $\gamma$.}
    \label{emean}
\end{figure}

Above a certain energy, the light produced by the EAS development
will provide a strong enough background to completely overwhelm
the D\v{C} light intensity, thereby making reliable measurements
impossible. The D\v{C} light can therefore only be determined
over a limited energy window.  The lower energy threshold is
defined by the threshold for \cer emission, set by $\gamma_0$
(Figure \ref{emean}), while the upper energy threshold is set as
the energy at which the secondary EAS \cer light overwhelms the
D\v{C} light.

For heavy nuclei the lower total particle energy threshold for the
observation window, $E_l$, can be described by $E_l=k_1 Z$ where
$k_1$ is a constant. The upper energy limit for D\v{C}
observation, $E_h$, occurs when the D\v{C} light level is
essentially equal to the background light level. Since D\v{C}
light emission $N_{D\check{C}}$ is proportional to $Z^2$, we can
express this as $N_{D\check{C}}= k_2 Z^2.$ The EAS \cer light
$N_S$ is proportional to E, $N_s=k_3E.$ Combining these last two
equations gives an expression for the upper energy observation
limit:

$$E_h=(k_2/k_3)Z^2$$

Hence the relative width of the observation window $\Delta
E/E_l=(E_h-E_l)/E_l$ is

$$\Delta E/E_l = (k_2/k_3 k_1)Z-1$$

\begin{figure}
  \centering
  \includegraphics[height=4.0in]{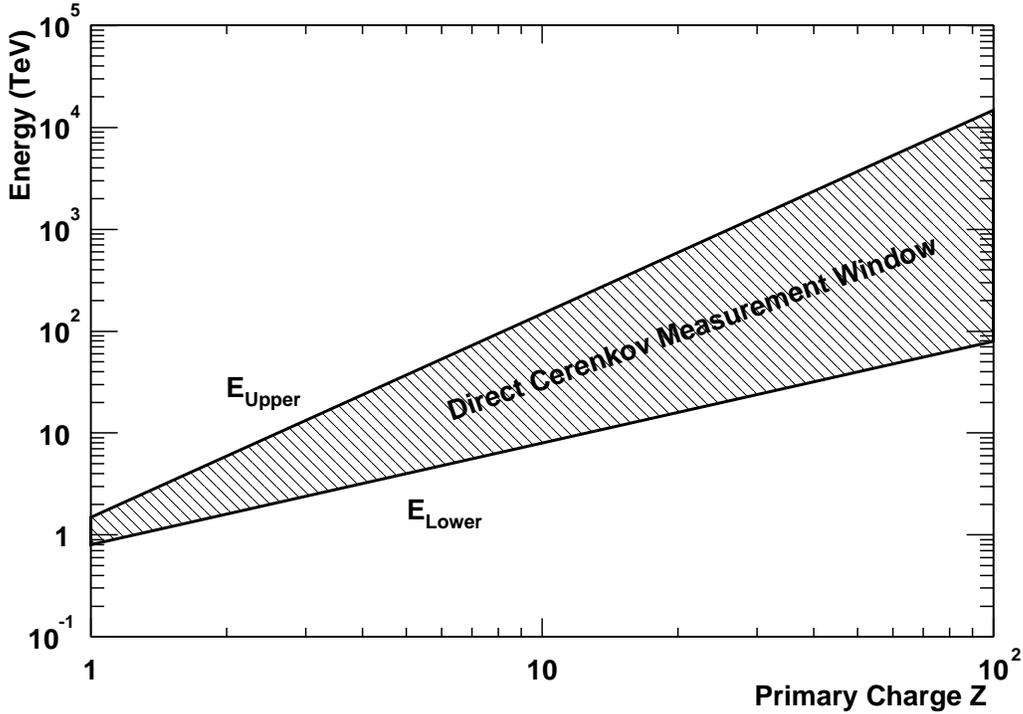}
  \caption{Upper and lower threshold for detection of D\v{C} light in
           cosmic ray air showers. The lower threshold is due to
           the \cer photon emission threshold. The upper threshold is where
           secondary light density from the EAS has a strength equal to the
           D\v{C} light density. Vertical Axis:
           Primary Energy (TeV). Horizontal Axis: Primary particle charge
           Z.} \label{erange}
\end{figure}

This window expands like $\sim Z$ for heavy nuclei. This arises
because the D\v{C} light scales like $Z^2$, whereas the background
level scales like $Z$.  The results of numerical
simulations give similar results to this simple estimate. Figure
\ref{erange} shows the upper and lower energy limits for D\v{C}
observation as a function of primary particle charge $Z$. A D\v{C}
measurement of Iron nuclei ($Z=26$) is possible into the
cosmic-ray ``knee'' region, around $10^3$ TeV.

\begin{figure}
  \centering
  \includegraphics[height=6.0in]{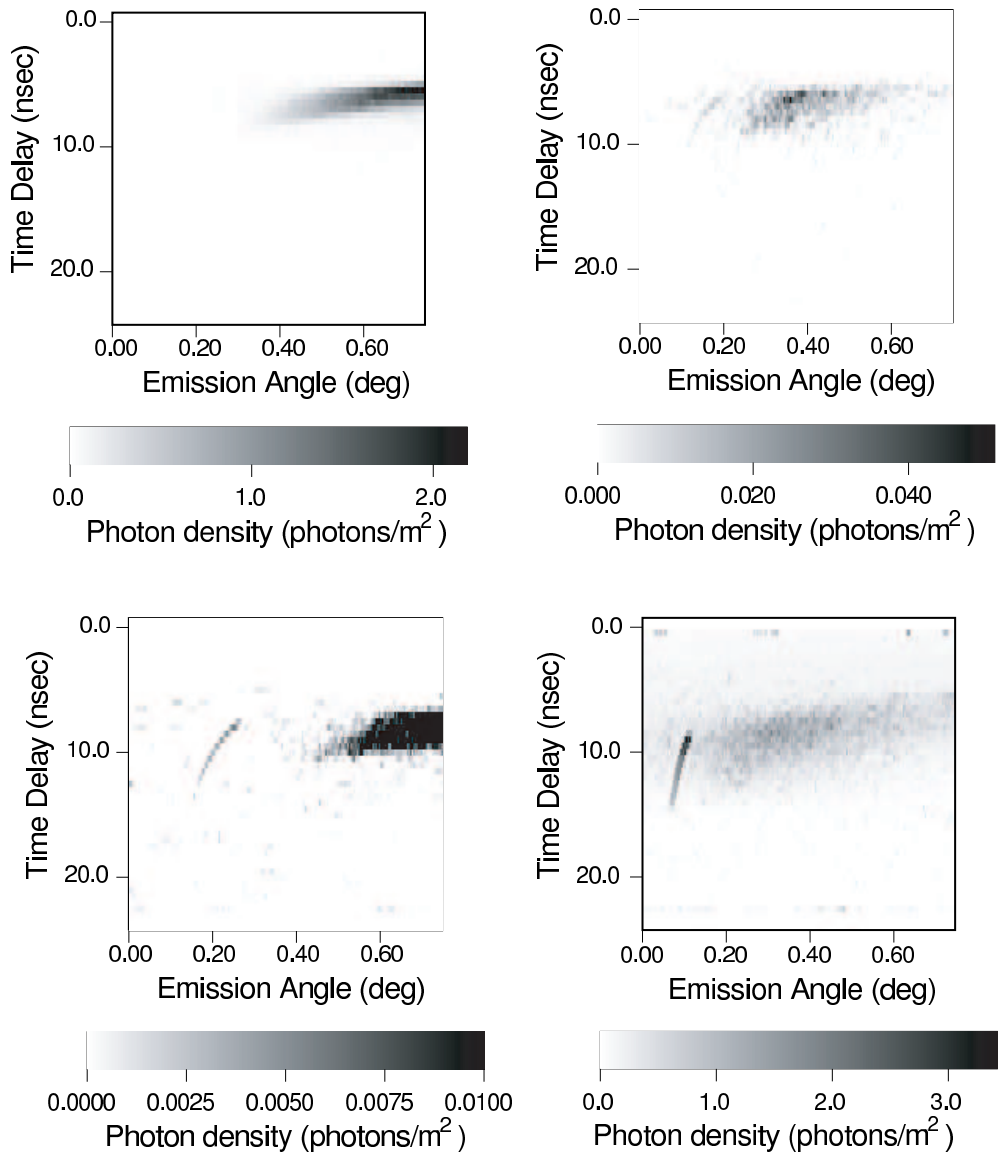}
  \caption{Simulated D\v{C} and EAS \cer light emitted from a single
            particle. The upper left hand panel shows a 10 TeV $\gamma$ at
            vertical incidence. The upper right panel is 100 GeV electron at
            vertical incidence. The lower left panel shows a 250 GeV proton at
            vertical incidence, and the lower right panel shows a 200 TeV
            $Z=50$ nucleus with a trajectory at $45^\circ$ to the zenith. The
            \cer light is observed at a radius between $67-94$ m (mean radius 80 m) from the
            shower axis. The scales below each panel give the photon
            intensities. The vertical axis is the time delay of the arriving
            photons as discussed in the text. The horizontal axis is the
            arrival angle of the photons with respect to the vertical at the observing
            site. The scales below each panel give the photon intensities.}
  \label{new4}
\end{figure}

The separation between the D\v{C} light and the secondary \cer
light persists at relatively large cosmic ray zenith angles.  The
lower right panel of Figure \ref{new4} shows the image from a
theoretical $Z=50$ nucleus at $45^\circ$ to the zenith. Since both
the D\v{C} light radius and the size of the EAS \cer disk grow
proportionally with the distance to the observation level, the
separation between these two light emission regions persists. An
advantage in large angle observation is that the light pool size
grows geometrically with zenith angle, thereby increasing the
effective detection area for cosmic ray observation. This may
prove important for increasing the detector collection area at the highest energies
(large Z) where cosmic ray particle fluxes are expected to be low.

\begin{figure}
 \centering
 \includegraphics[width=5.0in]{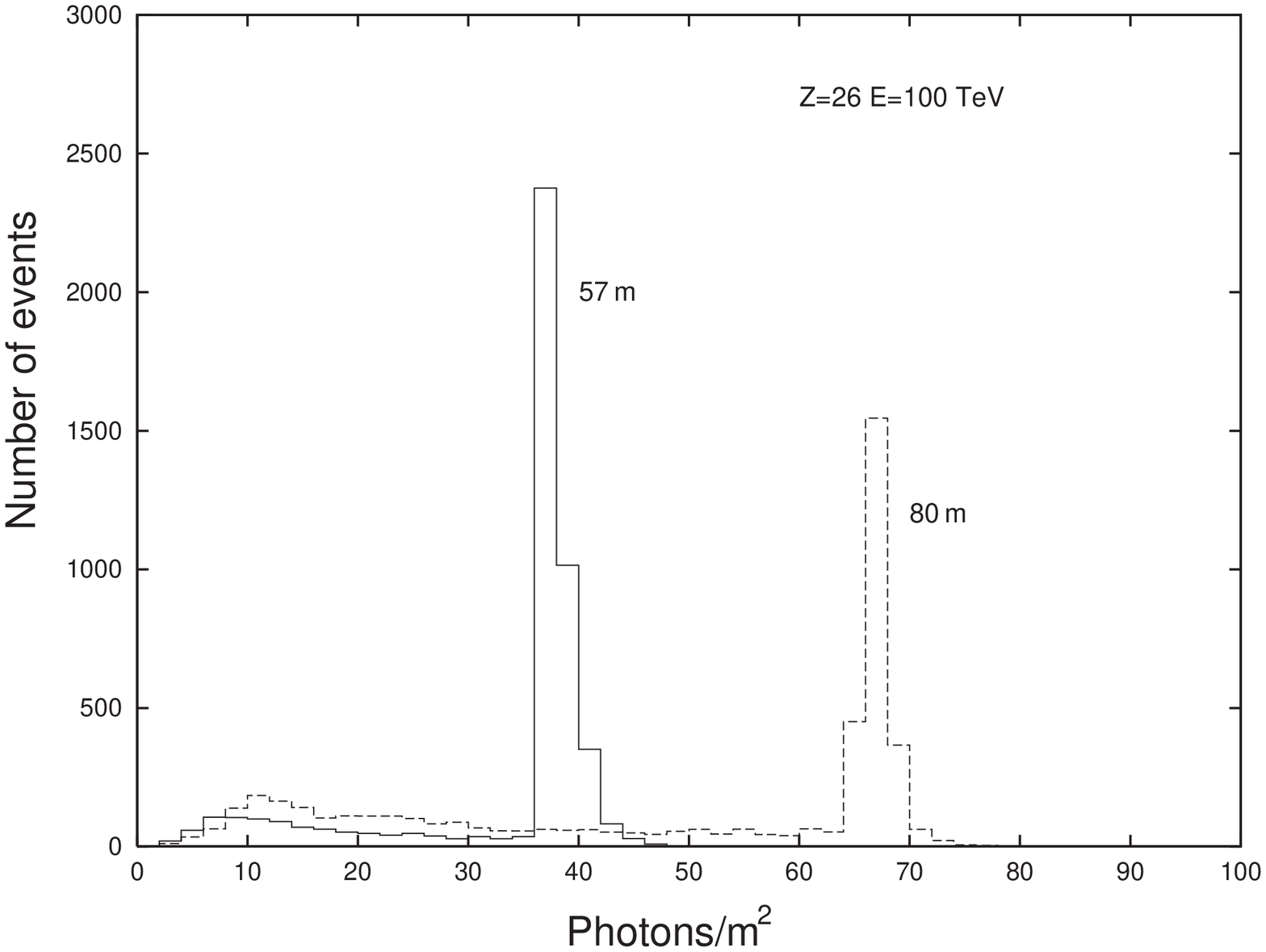}
  \caption{Histogram of D\v{C} photon yield observed at mean radii of 57 m and 80 m
           from shower core for a large number of 100 TeV iron
           nuclei (Z=26). Horizontal Axis: Photon density
           (photons/m$^2$). Vertical axis: Number of Events}
  \label{flucts}
\end{figure}

\begin{figure}
  \centering
  \includegraphics[width=3.5in]{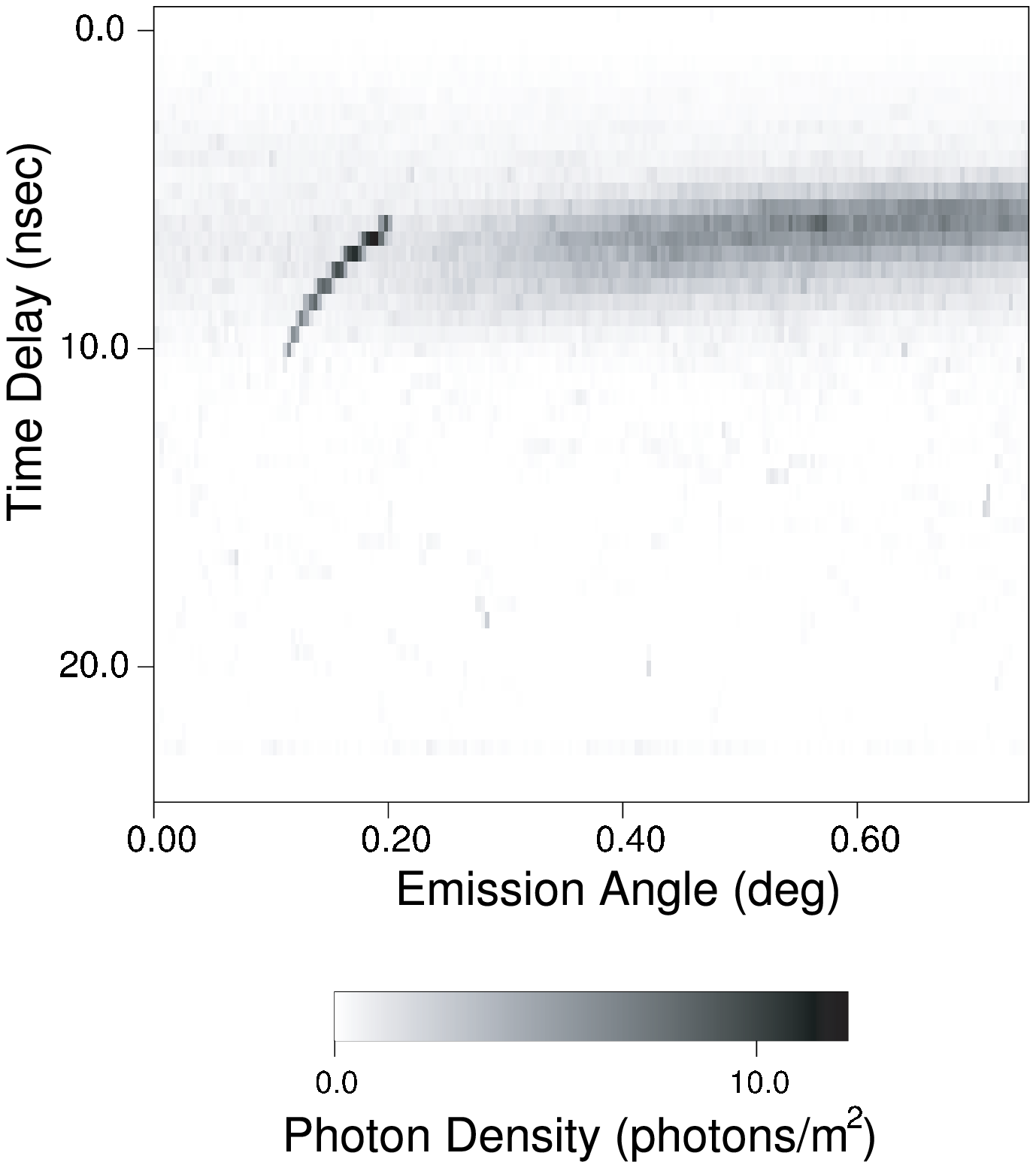}

   \caption{Direct and EAS \cer light emitted from 100 TeV $Z=26$
            nucleus with a vertical trajectory. The \cer light is
            observed at a mean radius of 80 m from the shower axis.
            Multiple scattering of the \cer light due to Rayleigh
            and Mie scattering from a wind-driven aerosol model of
            the atmosphere is included. The vertical axis is the time delay of the arriving
            photons as discussed in the text. The horizontal axis is the
            arrival angle of the photons with respect to the vertical at the observing
            site. The scales below each panel give the photon intensities.  }
  \label{fig:Mie}
\end{figure}

\section{Resolution Considerations}\label{sec:Flucts}
Above threshold, the D\v{C} light emission is independent of
primary energy, making it an ideal measure of the primary charge.
However, due to various background, detector resolution, and fluctuation processes,
 what one actually measures is not only the D\v{C} light emission, but rather the
integrated D\v{C} light yield plus all background across the
length of the primary cosmic track. We describe  each of these
effects and their Z dependence, and present a composite plot of
the predicted charge resolution  effects as a function of the
primary charge Z.

\subsection{Background Light Levels }
Background light sources which may degrade  the charge resolution
of this technique include fluctuations in  the night sky
background level, fluctuations in the secondary \cer light
emitted by the extensive air shower, and secondary light which is
scattered by the atmosphere to the same direction and arrival time
as the D\v{C} light signal.

\subsubsection{Night Sky Background}
The overall background expected by fluctuations in the night-sky light level can be
determined from previous measurements at dark sites. The standard
background value quoted is $2 \times 10^{12}$ photons/(m$^2$ sec
sr) over the range $300-600$ nm. This translates to an overall
rms intensity scale near 0.01 photons/m$^2$ for the angle/time bin sizes
used in Figure \ref{fig:Mie}, substantially smaller than
the typical D\v{C} signal. This quantity is independent of
the primary charge Z. Since the D\v{C} signal increases as $Z^2$,
the night sky background contribution to the
charge resolution $\Delta Z/Z$ decreases like  $1/Z^2$.

\subsubsection{Secondary \cer Background}

Since the threshold energy for observing \DC light increases
linearly with increasing Z, the secondary \cer light background,
which is proportional to primary energy, also increases linearly.
In principle, since the primary cosmic ray energy is measured
from the secondary \cer light, the amount of secondary background
light in the \DC pixel bins could be estimated. This could be
subtracted out to yield a pure \DC light measurement. However,
Poisson variations in the secondary \cer background will generate
fluctuations in the background-subtracted signal, thereby
limiting the \DC light measurement resolution. Consequently, the
contribution of the secondary \cer background to the charge
resolution has the form $$\Delta Z/Z \propto {{\sqrt Z} \over
Z^2}=Z^{-1.5}$$ In principle, optimal charge resolution is
obtained by matching the detector pixel size and time resolution
to the inherent width of the \DC light emission. In our
simulation plots, we have used time bins of 500 ps and angular
bins of 0.00375$^\circ$, close to the optimal values. However, a
conservative estimate of the magnitude of the secondary light
yield has been determined from the simulations using a coarse
0.2$^\circ$ by 6 ns square integration window about the D\v{C}
light yield position for a typical shower, in the absence of the
D\v{C} light signal. This larger bin size is typical of existing
state-of-the-art ground based \cer observatories.

\subsubsection{Atmospheric Scattering} The emitted D\v{C} light can
possibly be scattered in the lower regions of the atmosphere. Mie
and Rayleigh scattering are effects which scatter light from the
EAS \cer component into the D\v{C} light beam and also in to the
gap in the time versus angle image (see Figure \ref{new2}) which
separates the EAS emission and the D\v{C} light. Rayleigh
scattering by air molecules is essentially constant with respect
to atmospheric conditions. The Rayleigh process scatters light to
large angles (typically 90$^\circ$) with respect to the original
light beam direction, and therefore should provide very little
effect on the separation between D\v{C} light and EAS \cer
light.  Its main result is to increase the general background
light level in every pixel rather than to favor a particular
pixel. Less than 5\% of the total \cer light is scattered from
the narrow image, essentially isotropically. The general increase
in background light level due to Rayleigh scattering is
negligible.

Mie scattering is scattering from small suspended particles
(aerosols) in the atmosphere, and is highly dependent on the
aerosol composition, particle size distribution and height
distribution.  The Mie process acts to deflect the original photon
beam out to angles of 5--10$^\circ$ off-axis, which is
substantially more narrow than the characteristic Rayleigh angle.
Because the Mie-scattered light is deflected to relatively narrow
angles with respect to the original beam it represents a more
serious potential to degrade the performance of the D\v{C}
technique. The amount of light scattered from the original beam
depends strongly on the aerosol concentration. There are times
when absolutely no light is Mie scattered (``molecular"
atmosphere), and there are times when the atmosphere may be
essentially opaque due to rain, dust storms, pollution, etc. If
one is careful to select observation sites which have a large
number of purely molecular atmospheric condition at night, one
may be able to avoid this problem altogether.

For example, the HiRes Fly's Eye observatory, located at the
Dugway Proving Grounds, Utah, has a large fraction of its
observing time ($\sim 90$\%) with viewing conditions similar to
or better than the ``standard desert atmosphere''
\cite{Sokolsky96}. In this wind-driven aerosol model, typically
$5-20$\% of the total light beam is diverted uniformly out to
angles of $5-10^\circ$ off axis. Because the scattered beam is
 wider than the original \cer beam width, the general effect is to
slightly elevate the background light levels in all pixels rather
than in just the pixels relevant for the D\v{C} light
measurement. Using small pixels, the effect of the Mie scattered
light is substantially reduced. Figure \ref{fig:Mie} shows the
level of background light, including atmospheric scattering, for
a 100 TeV iron nucleus event. Clearly the D\v{C} emission remains
prominent. At this level, the Mie scattering will likely affect
the D\v{C} emission from the lighter nuclei ($Z<6$). Heavier
nuclei will be unaffected as their D\v{C} light is much stronger
than the Mie scattered light from the EAS.

For purposes of this paper, the resolution limitations due to
atmospheric scattering of the secondary \cer light is included in
the secondary \cer background calculation (Section 4.1.2).

\subsection{Detector Resolution Effects}
The detector-induced limitations to the charge resolution include
the core position resolution, angular reconstruction resolution,
and the error in the measured signal due to fluctuations in the
D\v{C} signal photo-electron statistics (which depends upon the
camera Quantum Efficiency ($QE$) and the Mirror area ($A$)). In
the below calculations we assume $A=100$ m$^2$  and $QE=25\%$.

\subsubsection{Angular Reconstruction Error}
For typical gamma-ray observatories like VERITAS\cite{VERITAS}
and HESS\cite{HESS}, a primary trajectory angular reconstruction
error of $\approx 0.1^\circ$ is expected. In these simulations,
we assume the high-resolution cosmic ray detector has a similar
angular reconstruction error. The charge resolution error is then
computed by comparing average D\v{C} light yields at a mean
radius  of 80 m for an ensemble of vertical simulated showers with
another ensemble of showers with identical charge and energy, but
with a primary zenith angle trajectory of 0.1$^\circ$. The
resulting error is independent of Z, and very small ($\Delta Z/Z
\approx 0.34\%$).

\subsubsection{Core Position Error}
Given a gamma-ray observatory like VERITAS\cite{VERITAS} with an
array of 10 m diameter primary mirrors, the core position can
only be localized to approximately 5 meters. Assuming a 5 meter
core resolution error, we examine the change in the D\v{C} light
yield as a function of distance from the shower core using the
Monte Carlo simulation. Figure \ref{flucts} shows the results of
simulations of several thousand vertically incident 100 TeV iron
nuclei viewed in a $0.2^\circ$ angular by 6 ns delay-time bin
centered on the D\v{C} emission region. The dashed histogram
shows the D\v{C} light observed in an annulus of $67-94$ m
(mean radius 80 m) of the particle path, the solid histogram
shows the D\v{C} light observed within an annulus of $47-67$ m
(mean radius 57 m). Clearly both of these observations have well
defined peaks. Simulations of different Z nuclei show virtually
the same ratio of  the peak D\v{C} light yield at mean radii of
57 m to 80 m. Using a linear interpolation of the light yield
variation with distance to the shower core, one derives a charge
resolution error of 3.1\% due to core position uncertainty. This
charge resolution error is independent of the primary charge $Z$.

\subsubsection{Photon Statistics}
Fluctuations in the photon statistics play  an  important role
for low D\v{C} light emission levels (at small Z).  Since the
signal is proportional to $Z^2$, the  charge resolution scales
like
$$\Delta Z/Z \propto {\sqrt{Z^2} \over Z^2} = {1/Z}.$$ The magnitude of signal is
determined from the Monte Carlo Simulation, assuming the above
specified values for QE and A.

\subsection{Hadronic Interaction Fluctuations}

Because the primary particle will eventually suffer hadronic
interaction, the length of the D\v{C} light emitting region
changes from event to event with the fluctuation in the depth of
the first interaction.  This leads to a subsequent fluctuation in
the D\v{C} light yield.  If the interaction point can be
identified in the track image (appearing as a sudden drop in
emission rate in a sufficiently finely pixelation detector), the
integrated yield could be normalized to path length.  However,
such an analysis in reality is complicated by a finite detector
resolution, photon statistics, and the fact that the nucleus does
not usually suffer a catastrophic collision in the first
interaction. Typically, a nucleus of charge $Z$ will break up
into a charge $Z-2$ nucleus and a Helium ($Z=2$) nucleus, and
therefore continue to emit D\v{C} light at only a modestly reduced
rate until it suffers further collisions. Fluctuations in this
fragmentation process therefore affect the integrated D\v{C}
light yield in a fashion for which there may be no simple
correction available.

\begin{figure}
  \centering
  \includegraphics[width=5.0in]{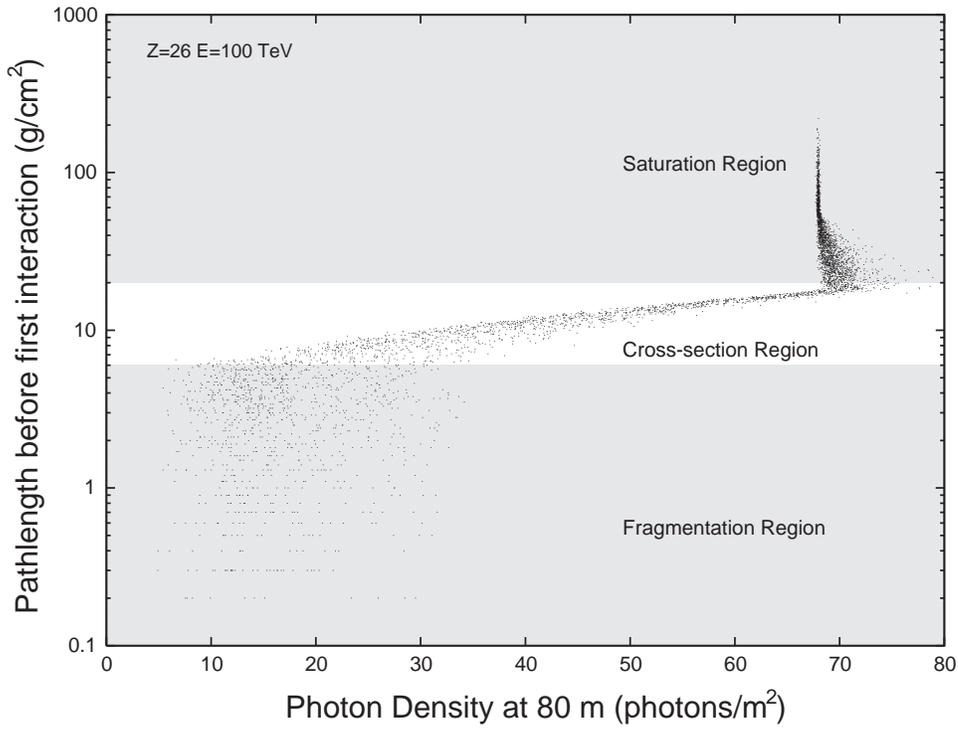}
   \caption{Correlation of D\v{C} emission with pathlength in atmosphere
    before first interaction for 100 TeV Iron primary ($Z=26$).
    Horizontal Axis: D\v{C} light intensity at $67-94$ m (mean radius 80 m) from shower core
    (photons/m$^2$). Vertical Axis: Pathlength in atmosphere before first
    interaction (g/cm$^2$).}.
  \label{scatter}
\end{figure}

The fluctuation effects are best illustrated by examining
characteristics of a large ensemble of simulated showers. A
scatter plot (Figure \ref{scatter} ) illustrates the distribution
of integrated \DC light measured between $67-94$ m from the shower
core as a function of the depth of the first nuclear interaction
of the primary. The scatter plot has essentially three regions of
interest, depending upon the depth of the first interaction.

For cosmic rays which interact less than 7 g/cm$^2$  deep (\ie,
at altitudes above $\sim 40$ km), there are large fluctuations in
the \DC light yield. Even nuclei with the same first interaction
depth suffer from large fluctuations. The \DC light yield
fluctuation is due mainly to the details on how the first
interaction proceeds. If the initial interaction yields a large
number of low Z fragments, the light yield is low. If the
interaction yields only 1 or 2 low Z fragments and the original
primary only loses a small amount of charge, then the \DC light
yield is high. This region (interaction depth $< 7 $ g/cm$^2$)
may be referred to as the `fragmentation region' as the magnitude
of the \DC light yield in this region is dominated by the details
of the fragmentation process. The fragmentation region is
illustrated in lower region of Figure \ref{scatter}.

For cosmic rays which interact between $7-18$ g/cm$^2$ in the
atmosphere, there is a tight linear relationship between depth of
first interaction and the \DC light yield. In essence, as the
particle penetrates deeper into the atmosphere, the total amount
of \DC light emitted by the full nucleus begins to dominate any
\DC light fluctuations generated by fluctuations in the
fragmentation process. The \DC light yield can be though of
purely geometrically, as the length of the path traveled between
$7-18$ g/cm$^2$ before the first interaction in this region. This
region may be called the `cross-section' region as the \DC light
yield in this region is dominated by the value of the inelastic
cross-section. The cross-section region is illustrated in middle
region of Figure \ref{scatter}.

For first interactions occurring deeper than 18 g/cm$^2$ (\ie,
altitudes below $\sim 30$ km), the light yield is essentially
independent of the interaction depth. This phenomena arises
because the \DC light is emitted at a specific emission angle.
The emission angle, when coupled with the height of emission,
defines the radius at which the \DC light will be observed. When
the particle passes beyond 18 g/cm$^2$ without interacting, all
the \DC light emitted beyond this depth falls outside the $67-94$
m radial bin used in this plot. Consequently, once the nucleus
has passed through 18 g/cm$^2$ without interacting, all light that
can be emitted into the $67-94$ m annulus has been emitted,
independent of where the subsequent first interaction takes
place. This region may be referred to as the `saturation region',
as the \DC light yield is essentially saturated to its maximum
value. The saturation region is illustrated in upper region of
Figure \ref{scatter}.

If a detection annulus closer to the particle path is used, the
D\v{C} collected will be emitted higher in the atmosphere, as
shown in Figure \ref{fig:Lines}. The altitude or grammage
boundaries between the fragmentation, cross-section, and
saturation regions are therefore functions of the radial distance
to the shower core. In this case a particle can interact higher
in the atmosphere (still below 40 km), existing in the
cross-section region for the outer radius (80 m)  but still be in
the saturation region for this inner radius (57 m).  In addition,
the overall light yield in these regions decrease as one
decreases the observation radius.

Since particles can exist in the saturation region for one radius
and be in the cross-section region for a larger radius, it is
possible to extract additional information by examining the ratio
of \DC light emitted in different radial bins. For instance, for
a particle in the saturation region at 57 m but in the
cross-section region at 80 m, the ratio of \DC light measured at
the 80 m distance to the \DC light measured at the 57 m distance
is strongly correlated with the depth of the first interaction.
Consequently, a model-independent measurement of the inelastic
nuclear cross-section is possible, even with moderately coarse
camera pixels. Determination of size of the fluctuations in the
fragmentation process may also be possible using three radial
measurements: an outer radius measurement in the fragmentation
region, a middle radius measurement in the cross-section region,
and an inner radius measurement in the saturation region.

\begin{figure}
  \centering

  \includegraphics[width=5.0in]{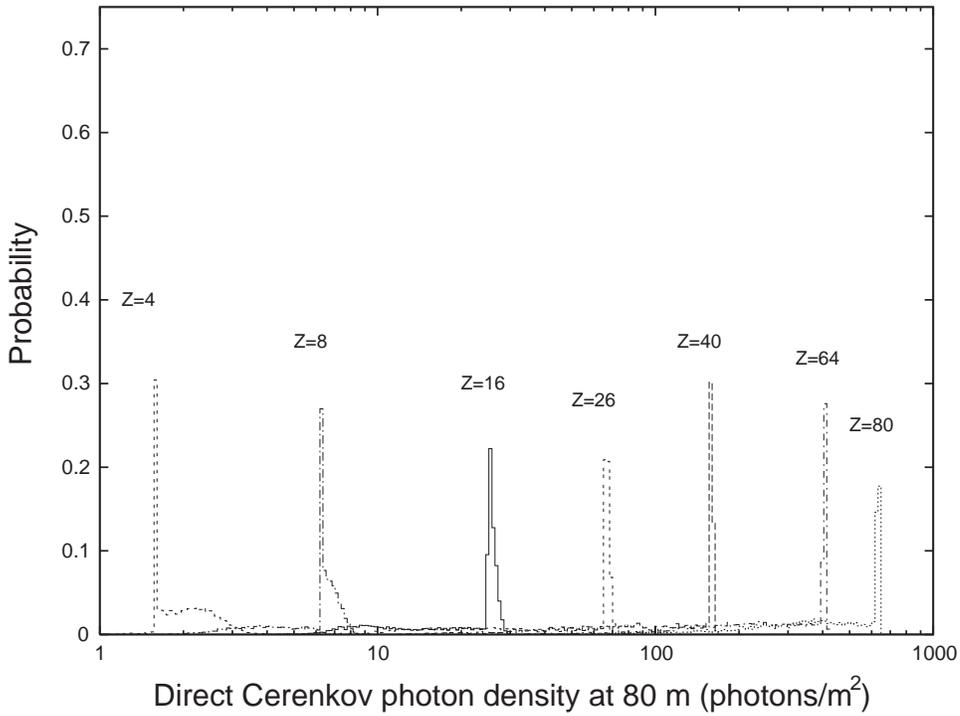}
   \caption{Distribution of D\v{C} light emitted
    at mean radius 80 m from shower core
    for several thousand simulated showers at each primary Z.
     Horizontal Axis: D\v{C} light intensity (photons/$m^2$). Vertical Axis:
    Probability (1.0 =100\%)}.
  \label{Zdist}
\end{figure}

An important concept is that all possible interaction fluctuations
can only result in a  {\em decrease} in the observed \DC light level.
This makes \DC light distribution in Figure \ref{Zdist} asymmetric
 at the {\em lower} \DC light intensity level. For example,
 although $Z=64$ nucleus might occasionally yield a very low light
\DC signal comparable to a $Z=26$ or $Z=40$ nucleus, a {\em
smaller charge} (\eg, $Z=40$) nucleus cannot emit the \DC light
intensity of a $Z=64$ nucleus. Consequently interaction
fluctuations cannot result in an overestimation of the particle
charge, it can only underestimate it.

The interaction fluctuation contribution to the overall
resolution is determined from the  width of each narrow peak
distribution in Figure \ref{Zdist}.  The resulting Z dependence
of the charge resolution is found to be well described by a power
law $$\Delta Z/Z \propto Z^{-0.73}.$$

\subsection{Overall Resolution}
\begin{figure}
 \centering
 \includegraphics[width=5.0in]{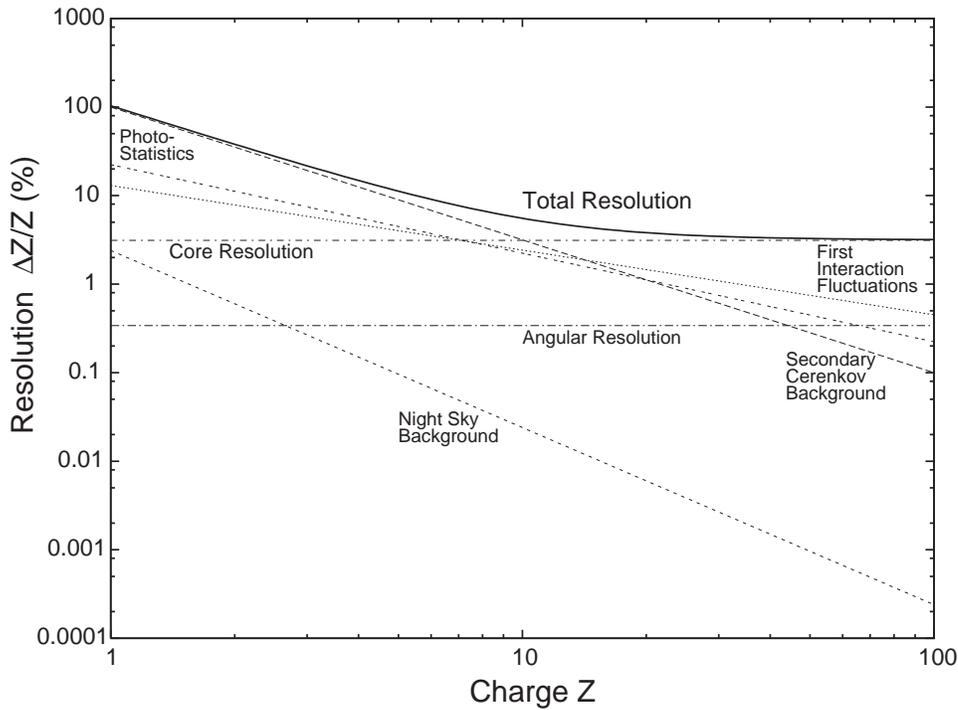}
  \caption{The expected charge resolution $\Delta Z/Z$ for a
  detector of effective area 100 m$^2$ and core position resolution 5 m. Horizontal Axis:
Primary Charge Z. Vertical Axis: Charge Resolution $\Delta Z/Z (\%)$ }
  \label{zres}
\end{figure}

The charge resolution expected from the \DC technique is limited
by the combination of the various effects described above. In
order to be conservative, we have assumed a detection scheme with
an effective collecting area of 100 m$^2$, a core location
capability of 5 m, a time resolution of 6 ns, and an angular
pixel size of 0.2$^\circ$. Figure \ref{zres} shows the charge
resolution expected resulting from these effects as a function of
charge $Z$.  For low charges the resolution is dominated by
secondary \cer light from the EAS. At higher $Z$ the core
resolution provides the charge resolution limitation. The overall
resolution is calculated to be $\Delta Z/Z \sim 5\% $ for $Z>10$,
essentially independent of charge.

In principle, improved charge resolution could be obtained by
matching the detector pixel size and time resolution to the
inherent width of the \DC light emission (about 500 ps timing and
0.00375$^\circ$ wide pixels). Since the resolution function for
high Z is dominated by core position error contribution, this
would not improve the resolution at large Z. However, it could
substantially improve the resolution for $Z<10$ as it would
reduce the effects of the secondary \cer light in the charge
resolution.

\section{Detector Design and Applications}\label{sec:Straw}

A possible design for a D\v{C} imaging observatory can be
constructed from the simulation work results. Such an observatory
comprises a number of large ($>$10 m diameter) front surface
reflecting dishes in fixed mounts. The mounts point vertically
for low energy composition measurements, but are adjustable to
view at large zenith angles to increase detection area for
measuring the higher energy cosmic ray heavy nuclei. The dishes
are separated by a distance of $\sim$ 80 m for sea level
operation, placed on a hexagonal grid pattern.

Each dish has an isochronous surface in order to preserve the
incoming wavefront timing to approximately 2 ns or better. This
timing requirement provides additional separation between the
D\v{C} light and the secondary \cer light as shown in Figure
\ref{new2}. A highly pixelated optical 
camera built from small, fast photomultiplier tubes
is used at the focal plane to image the \cer light from the
cosmic ray shower. In order to have the best resolution and
signal-to-noise ratio for the D\v{C} light, small pixels
(typically $0.01^\circ$) are preferred. However, the D\v{C} light
measurements can be made with large pixels, up to
$0.1-0.15^\circ$. A field of view of $5-10$ degrees is needed to
provide a sufficient collection aperture. For a $10^\circ \times
10^\circ$ field of view, and a fiducial area defined between
$50-115$ m of a detector, each dish has an effective collection
aperture of $\sim 1000$ m$^2$sr.

Light striking the individual pixels is read out by a fast FADC
system, with 500 MHz sampling rate or faster, enabling an image
to be constructed in time versus/angle like those shown in Figure
\ref{new2}. A key advantage of the D\v{C} method over the
observations of \cer light high in the atmosphere is that the
triggering of the system is relatively simple. Since a significant
part of the EAS will develop in the field of view of the
telescopes, this large light signal can be used to provide a
simple trigger scheme. A \cer light signal produced by an EAS with
energies $>$ 10 TeV can be reliably discriminated from the night
sky by simple logic on the camera pixels for a 10 m mirror size.
Using individual-tube constant fraction discriminators, hit
patterns from multiple pixels are combined using fast logic
tables to look for sequences of pixels with characteristics
similar to a \cer image. Once a mirror is triggered, it retrieves
a history of photon time slices from the FADC system for about
$20-50$ nanoseconds around the trigger time to look for the
delayed D\v{C} light signal. Neighboring telescopes in the array
would also be read out to look for coincidence measurement of the
same \cer image from different observation angles. The data from
multiple telescopes is combined to determine the shower geometry,
energy, and D\v{C} light content. Importantly, events can be
required to contain consistent amounts of D\v{C} light intensity
and location in multiple telescopes. This procedure can reject
essentially all light signal contamination from local sources,
for example local muons, and verify the level of D\v{C} emission
from independent measurements.

The total detection aperture is not only affected by the physical
size of the obervatory and the field of view of each camera; it
is also affected by additional constraints needed to observd  the D\v{C},
 light such as requiring the shower core to strike in an annulus between 50-115 m from the
individual detectors. Because we are using an array of multiple telescopes, in
general the shower core will fall within an acceptable distance from more
than one telescope over most of the area surrounding the array. 
Additional considerations, such as field of view, also restricts
the detection aperture. These considerations are similar to those
imposed on typical ground-based gamma ray observatories (e.g. VERITAS), and consequently
we expect a similar detection aperture to these devices (i.e. several thousand m$^2$sr).
However, an accurate determination of the detector acceptance, including 
energy and charge dependence of the aperture, can only be
accomplished through detailed Monte Carlo simulation. This simulation must
include a detailed model of an assumed detector configuration (array spacing, mirror area,
pixel size, electronic triggering charateristics, etc). Monte Carlo calculations  for a typical 
strawman observatory configuration, modeled after the VERITAS observatory, are underway and
will be presented in a subsequent paper. 

We note that the next generation of imaging \cer telescope arrays,
including VERITAS\cite{VERITAS}, HESS\cite{HESS}, and CANGAROO\cite{CANGAROO}
 all begin to approach the
strawman design, and will likely be able to provide a first
measurement of D\v{C} light. An intriguing possibility is to use
the D\v{C} light to reject background in these experiments. The
ability to distinguish between the \cer signals produced
by electrons and gamma rays in the $10-500$ GeV energy range is
important for substantial improvement in the sensitivity of
ground-based gamma-ray astronomy. In this energy range, protons
and heavy nuclei do not produce secondary particles with
sufficient energies to generate substantial \cer light. The
majority of the background events for gamma ray instruments in
this region are produced by cosmic ray electrons, which also
generate pure electromagnetic air showers, with EAS \cer image
characteristics identical to a gamma ray primaries. The
capability to identify even a small fraction of the electron
events could make significant improvements in the sensitivity of
gamma-ray telescopes in this energy range. The upper left hand
panel of Figure \ref{new4} shows an angle/time image of a
simulated vertical gamma-ray shower at 10 TeV. The upper
left panel shows a vertical electron at 100 GeV. This second
event could be identified as an electron if the faint D\v{C}
image arc at the left of the panel is detected. The intensity in
this arc is $\sim$ 5 times the expected night sky background.

A similar background occurs in the 500 GeV $-10$ TeV range for protons
which transfer most of their primary energy to one or two $\pi^0$
secondaries in their first interaction, leaving little energy for
further hadronic interaction (\ie, generation of $\pi^\pm$ which
decay to muons). These types of events represents a substantial
amount of the residual cosmic ray background after applying
standard image analysis cuts to select gamma rays. Since these
proton-induced events are pre-selected to have image
characteristics similar to gamma rays, additional parameters must
be used to reject these events. The lower left hand panel in
Figure \ref{new4} shows an event in which the D\v{C} signal from
a proton has transferred most of its energy to an initial $\pi^0$
during the interaction. In comparison with the gamma-ray events
these type of events might possibly be rejected using
finely-grained timing and angular imaging.

\section{Conclusions}\label{sec:Conclusion}

We have discussed a new experimental technique which can
potentially yield excellent charge resolution measurements ($\Delta
Z/Z<5\%$ for $Z=26$) for ground-based observations of high energy
cosmic rays. The technique relies upon the observation of the
direct \cer light emitted by the primary nucleus before the first
nuclear interaction with the Earth's atmosphere. The experimental
technique works over an energy range in the TeV-PeV energy range,
with the width of the energy window growing like $Z$ for heavy
nuclei. A dedicated observatory could be built to observe cosmic
rays using his technique, while the next-generation ground-based
imaging \cer telescopes such as VERITAS, HESS, and CANGAROO will
begin to approach the sensitivity required to begin initial
observations.

The average intensity and fluctuations in the D\v{C} light yield
have been examined using Monte Carlo simulations. The yield
contains the expected $Z^2$ dependence with fluctuations that are
easily understandable in terms of the radial distance of
observation, variations in the depth of first hadronic
interaction, and the details of the subsequent fragmentation
process.  Indeed, D\v{C} light measurement appears to provide a
method to measure the primary charge independent of any hadronic
interaction or fragmentation model. This could potentially solve
the most difficult systematic problem in the ground-based studies
of TeV/PeV cosmic rays: the difficulty in distinguishing between
actual changes in composition and systematic shifts of nuclear
interaction models with energy. The high resolution nature of
this measurement might allow the detection of changes in the
propagation pathlength distribution at high energy. The
pathlength distribution at PeV energies is unknown, but has a
extremely strong effect on the predicted composition at the knee
of the all-particle spectrum \cite{Swordy:1995icrc}.

The D\v{C} light measurement technique could provide researchers
with a ``tagged particle beam'' of known energy and composition.
By examining D\v{C} light yield at various radial bins, one
may be able to extract nucleon-air cross-section information
as a function of primary mass and energy, independent of an
interaction model. Fluctuations in the D\v{C} light yield for
small interaction depths could provide information concerning the
fluctuations in the fragmentation processes of nucleus-air
interactions.

The D\v{C} light measurement technique may provide additional
sensitivity to ground-based imaging air \cer gamma-ray detector
for the rejection of the dominant electron background in the
$10-500$ GeV energy range, and residual proton background in the
$1-10$ TeV energy range. Such an improvement could potentially
increase the point source sensitivity of these telescopes by an
order of magnitude or more.

The unique properties of D\v{C} light (versus EAS light) may
allow the observation of exotic cosmic ray particles, such as
``strange quark matter'' (see, \eg,
\cite{Banerjee:1998wk,Banerjee:2000ye}) or magnetic monopoles
\cite{Wick:2000}.  These particles, which are thought to have
large effective charges, would provide an extremely strong \DC
signature due to the $Z^2$ dependence of the light yield.

A technique which combines the collection power and logistical
advantages of ground-based cosmic ray detectors with the
high-precision charge resolution of balloon or satellite-borne
experiments would be a valuable asset in the field of high-energy
astrophysics. Such a technique, if scientifically and financially
viable, has the potential to make great strides in the
determination of cosmic ray composition across the ``knee", and
subsequently, foster advances in the theory of cosmic ray
origins.  We believe that the development of imaging atmospheric
\cer telescopes has provided the means to target the \cer light
emitted directly from primary nuclei prior to their hadronic
interaction in the atmosphere. A direct \cer experiment,
implemented as an array of 10 m reflectors imaged by
high-resolution cameras with fast ADC systems may prove to be the
next step towards understanding the nearly century-old mystery of
cosmic rays.

\vspace{1ex}
{\bf Acknowledgements}
The authors express their appreciation for useful discussions with
members of the VERITAS collaboration during the preparation of this
paper. DBK acknowledges  computational support from the Utah
High Energy Astrophysics Institute.

\bibliographystyle{elsart-num}
\bibliography{DirectCerenkov}
\end{document}